\documentclass[aps,pra,twocolumn,amsmath,showpacs,floatfix]{revtex4}
\usepackage{bm}
\usepackage{graphicx}
\begin{document}
\setlength{\arraycolsep}{2pt}
\title{Quantum state engineering by a coherent superposition
of photon subtraction and addition}

\author{Su-Yong Lee }
\affiliation{Department of Physics, Texas A\&M University at Qatar,
 Education City, POBox 23874, Doha, Qatar}
\author{Hyunchul Nha}
\affiliation{Department of Physics, Texas A\&M University at Qatar,
 Education City, POBox 23874, Doha, Qatar}

\date{\today}

\begin{abstract}
We study a coherent superposition $t\hat{a}+r\hat{a}^\dag$ of field
annihilation and creation operator acting on continuous variable
systems and propose its application for quantum state engineering.
Specifically, it is investigated how the superposed operation
transforms a classical state to a nonclassical
one, together with emerging nonclassical effects. We also propose an
experimental scheme to implement this elementary coherent operation
and discuss its usefulness to produce an arbitrary superposition of
number states involving up to two photons.
\end{abstract}

\maketitle

\section{Introduction}
Manipulation of light field at the single-photon level provides a
crucial basis for many important applications in quantum information
science. In particular, two elementary operations on a single-mode
field, i.e. photon subtraction and addition represented by bosonic
annihilation and creation operators $\hat{a}$ and $\hat{a}^\dag$, respectively,
can be employed to transform a field state to a desired one \cite{Kim}.
For example, the photon subtraction transforms a Gaussian entangled
state (two-mode squeezed state) to a non-Gaussian entangled state
for a nonlocality test \cite{nha} and entanglement distillation
\cite{Takahashi}. The photon addition is known to create a
nonclassical state from any classical state (e.g. coherent and
thermal states) \cite{Agarwal}, and both of the photon-subtracted
\cite{Opatrny,Cochrane,Olivares,Kitagawa1} and the photon-added
squeezed states \cite{Li,Illuminati} were suggested to improve the
fidelity of continuous variable (CV) teleportation. The photon
subtraction \cite{Wenger} and the addition \cite{Zavatta} are now
practically realized in laboratory.

Recently, there appeared a proposal to implement a coherent
superposition $\hat{a}\hat{a}^\dag\pm \hat{a}^\dag \hat{a}$ of two
product operations, photon addition followed by subtraction
($\hat{a}\hat{a}^\dag$) and  photon subtraction followed by addition
($\hat{a}^\dag \hat{a}$) \cite{Kim1}. Kim {\it et al.} particularly
suggested an experimental scheme to prove the bosonic commutation
relation, $[\hat{a},\hat{a}^\dag]=1$ \cite{Kim1}, which was very
recently carried out using a thermal state of light field
\cite{Zavatta1}. A cavity-based scheme has also been proposed to
prove the bosonic commutation relation, exploiting interaction
between three atoms and a cavity field \cite{Park}. Based upon the
inteferometer setting of \cite{Kim1}, Fiur\'a\u sek suggested an
optical scheme to implement an arbitrary polynomial of photon-number
operators, e.g. noiseless amplifier \cite{amplifier} and Kerr nonlinearity, 
via a combination of multiple photon subtraction and
addition \cite{Fiurasek1}.

In this paper, we consider a coherent superposition of photonic
operations at a more elementary level, that is, the superposition of
photon subtraction and addition, $t\hat{a}+r\hat{a}^\dag$, and
investigate how it transforms a classical state to a nonclassical
one. In general, a coherent superposition of two distinct operations
may be created by erasing the ``which-path" information relevant to
the operations in an interferometer setting. Motivated by the
single-photon interferometer in \cite{Kim1}, we propose an
experimental scheme to implement the coherent operation
$t\hat{a}+r\hat{a}^\dag$ in an optical experiment and also show that it can be employed
together with displacement operators to engineer an arbitrary
quantum state in principle. As an example, we study the case of
generating an arbitrary superposition of number states involving up
to two photons. The superposition state
$C_0|0\rangle+C_1|1\rangle+C_2|2\rangle$ can be used for
quantum information processing, e.g., the nonlinear sign-shift(NS)
gate (a basic element of the nondeterministic CNOT gate)
\cite{Knill, Ralph}, and the optimal estimation of the loss
parameter of a bosonic channel \cite{Adesso}. There have been
several theoretical proposals to generate the superposition of
$|0\rangle$, $|1\rangle$, and $|2\rangle$, e.g., using repeated
photon subtractions with squeezing operations \cite{Fiurasek},
repeated photon additions with displacement operations \cite{Dakna},
and quantum scissors \cite{Pegg, Koniorczyk}. An experimental
realization was recently made conditioned on photodetections using a
parametric down-converter with two auxiliary weak coherent states
\cite{Bimbard}. We study our proposed scheme with experimental imperfections
considered and show that it can produce the superposition state with
high fidelity. 

This paper is organized as follows. In Sec. II, we study how the
coherent operation $t\hat{a}+r\hat{a}^\dag$ transforms a
classical (coherent or thermal) state to a nonclassical one in phase space with the degree of nonclassicality measured by negative volume (area) and by nonclassical depth \cite{CTLee}. In Sec. III, we
investigate observable nonclassical effects, squeezing and sub-Poissonian statistics, arising due to the coherent operation, and
propose an experimental scheme in Sec. IV to implement the operation $t\hat{a}+r\hat{a}^\dag$ in a single-photon inteferometer setting.
In Sec. V, we suggest the coherent operation combined with displacement operation for quantum state engineering and investigate the generation of arbitrary superposition state involving up to two photons with experimental imperfections included. The main results of this paper are summarized in Sec. VI.

\section{Wigner distribution}
We first examine how the coherent operation $t\hat{a}+r\hat{a}^\dag$
transforms a classical state to a nonclassical one in view of the
phase-space distribution. For a single-mode state $\rho$, its Wigner
distribution $W(\alpha)$ is generally given by the Fourier transform of the
characteristic function $C(\lambda)\equiv{\rm Tr}\{\rho
\hat{D}(\lambda)\}$, where $\hat{D}(\lambda)\equiv e^{\lambda
\hat{a}^\dag-\lambda^*\hat{a}}$ is the displacement operator. That
is, $W(\alpha)=\frac{1}{\pi^2}\int d^2\lambda
C(\lambda)e^{\alpha\lambda^*-\alpha^*\lambda}$.

(i) {\it Coherent states}:  Given a coherent state
$|\alpha_0\rangle$ as an initial state, the output state after the
superposition operation,
$|\Psi\rangle\sim(t\hat{a}+r\hat{a}^\dag)|\alpha_0\rangle$,
 possesses the Wigner
distribution
\begin{eqnarray}
W(\alpha)=\frac{
|t\alpha_0+r(2\alpha^{\ast}-\alpha^{\ast}_0)|^2-|r|^2}
{|r|^2+|t\alpha_0+r\alpha^{\ast}_0|^2}W_0(\alpha),
\end{eqnarray}
where $W_0(\alpha)\equiv\frac{2}{\pi}e^{-2|\alpha-\alpha_0|^2}$ is
the Wigner function of the initial coherent state $|\alpha_0\rangle$.
\begin{figure}[tbp]
\centerline{\scalebox{0.45}{\includegraphics[angle=270]{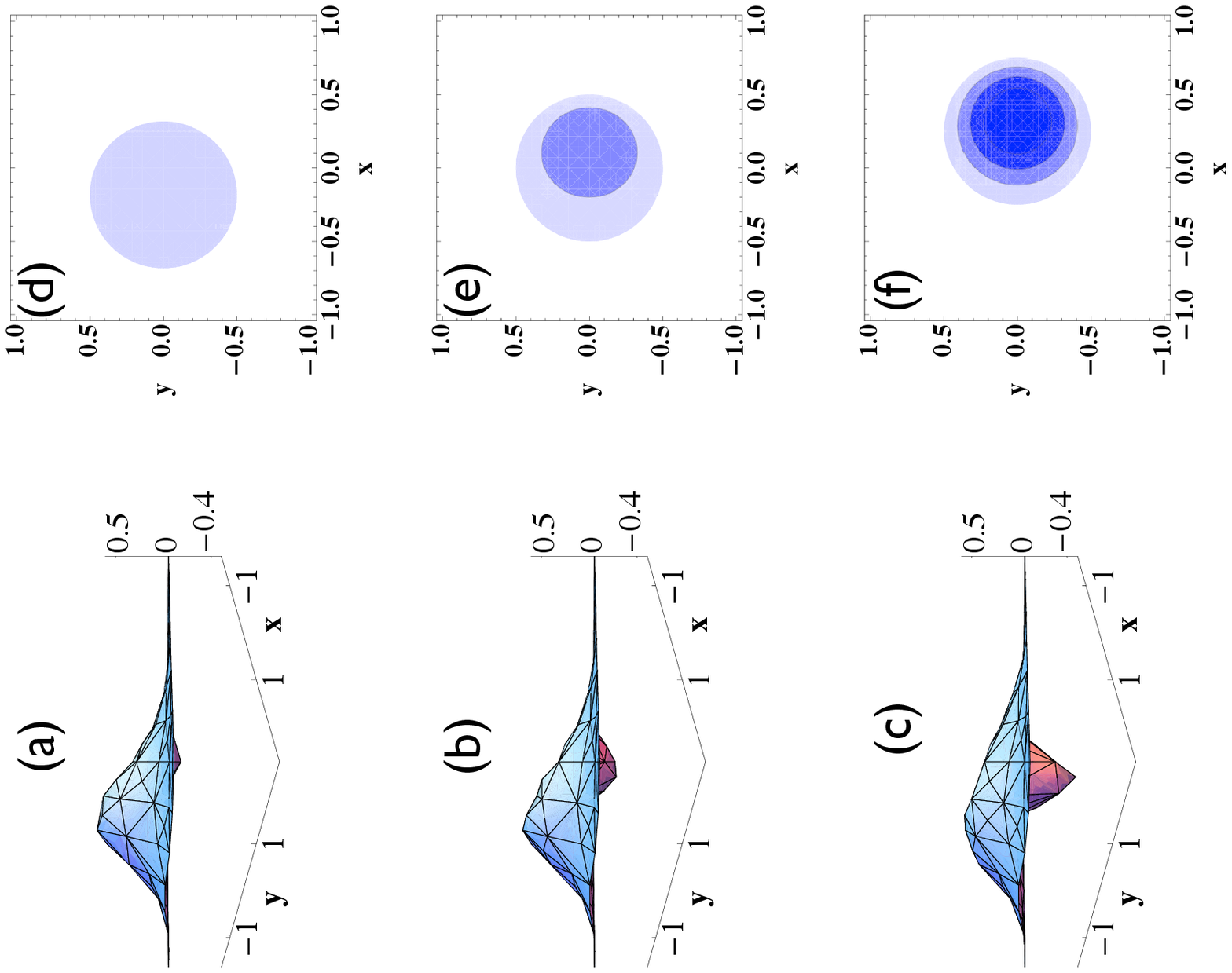}}}
\caption{Wigner distribution of the state
$(t\hat{a}+r\hat{a}^\dag)|\alpha_0\rangle$, where $|r|^2+|t|^2=1$, for $\alpha_0=0.5$ with (a) $r=1/2$, (b) $r=1/\sqrt{2}$
 and (c) $r=1$. (d), (e), and (f) are the contour plots corresponding to (a), (b), and (c), respectively, where
 only the negative regions are colored in blue. (d), (e), and (f) exhibit the
 same size $\pi/4$ of the negative area in phase space. }
\end{figure}
In Fig. 1 we show the Wigner distribution $W(\alpha\equiv x+iy)$ of the output state $(t\hat{a}+r\hat{a}^\dag)|\alpha_0\rangle$,
where $r$, $t$ and $\alpha_0$ are taken as real ($|r|^2+|t|^2=1$). In general, the negative dip of
the Wigner distribution increases with the ratio $r$ of the addition
operation $\hat{a}^+$ in the superposition $t\hat{a}+r\hat{a}^\dag$.
In order to measure the degree of nonclassicality for the output
state, we investigate both the negative area and the negative
volume in phase space.

{\it Negativity}---The negative region with $W(\alpha=x+iy)<0$
appears under the condition
\begin{eqnarray}
(x-C_1)^2+y^2< \frac{1}{4},
\end{eqnarray}
where $C_1=\frac{1}{2}(1-\frac{t}{r})\alpha_0$. It thus becomes a circle
of radius 1/2 regardless of $r$ [Fig. 1 (d), (e), and (f)], except the case of
$r=0$ which does not change a coherent state
($\hat{a}|\alpha_0\rangle=\alpha_0|\alpha_0\rangle$). Therefore, the
negative area is independent of the ratio $r$. However, the depth of
the negativity depends on $r$ [Fig. 1 (a), (b), and (c)],
which may be further quantified via the negative volume defined by
$V_N=\frac{1}{2}(\int d^2\alpha|W(\alpha,\alpha^{\ast})|-1)$
\cite{Kenfack}. We find that $V_N$ generally increases with $r$, as
shown in Fig. 2 (a). For a fixed $r$, the negative volume $V_N$ decreases with the initial
amplitude $\alpha_0$ and the extremal $V_N$ thus appears at
$\alpha_0=0$ (vacuum state), for which
$(t\hat{a}+r\hat{a}^\dag)|0\rangle=r|1\rangle$ (one-photon state) giving
$V_N=\frac{2}{\sqrt{e}}-1\approx0.2131$ regardless of $r$.

\begin{figure}[tbp]
\centerline{\scalebox{0.4}{\includegraphics[angle=270]{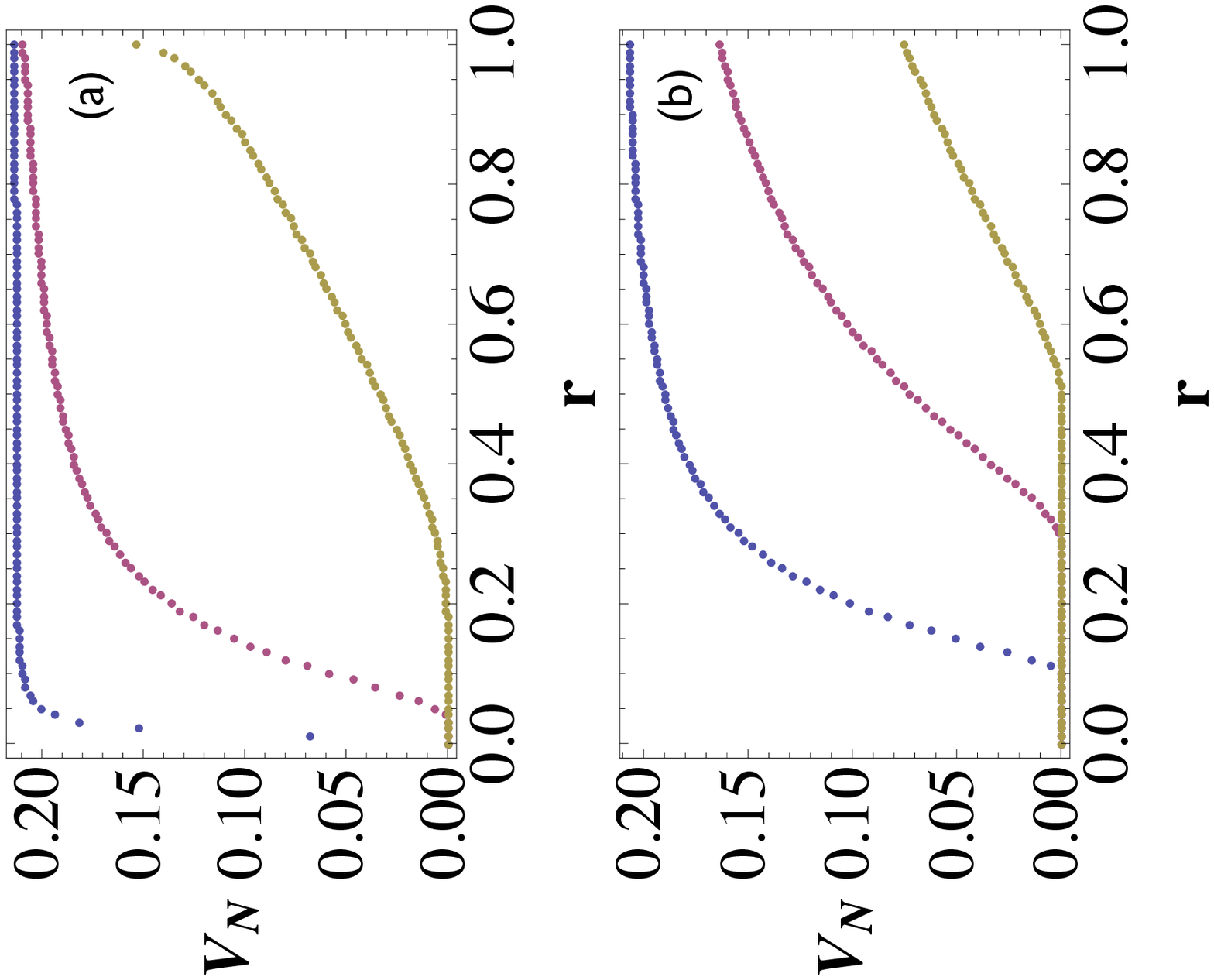}}}
\caption{Negative volume as a function of $r$ on applying the coherent operation $t\hat{a}+r\hat{a}^\dag$ (a) for a coherent-state input with $\alpha_0=$ 0.01, 0.1, and
0.5 (from upper to lower curves), and (b) for a thermal-state
input with $\overline{n}=$ 0.01, 0.1, and 0.5 (from upper to lower
curves) ($|r|^2+|t|^2=1$).}
\end{figure}

\begin{figure}
\centerline{\scalebox{0.4}{\includegraphics[angle=270]{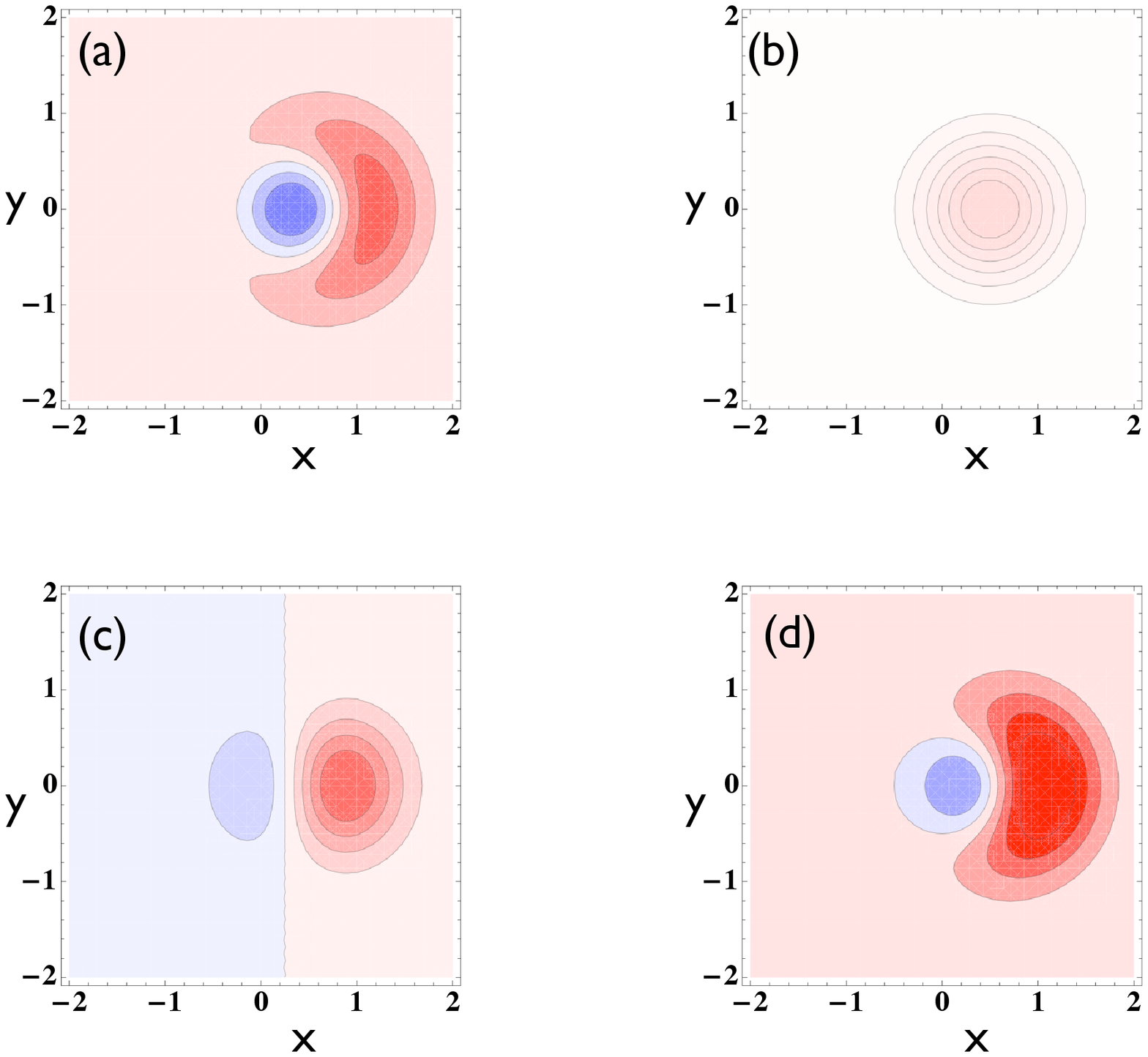}}}
\caption{Contour plot of the Wigner distribution after the coherent operation $t\hat{a}+r\hat{a}^\dag$ on the coherent-state input $\alpha_0=0.5$
with $r=1/\sqrt{2}$, for (a) $\rho_{+-}\equiv
\hat{a}^\dag\rho_0\hat{a}$, (b) $\rho_{-+}\equiv
\hat{a}\rho_0\hat{a}^\dag$, (c) $\rho_{++}+\rho_{--}\equiv
\hat{a}^\dag\rho_0\hat{a}^\dag+\hat{a}\rho_0\hat{a}$, and (d) entire
distribution.} \label{fig:fig3}
\end{figure}

Alternatively, one may quantify the degree of nonclassicality by the
nonclasical depth defined by C. T. Lee \cite{CTLee}. The Glauber-$P$
function, $P(\alpha)$, becomes positive when smoothed by a Gaussian
of sufficient width,
\begin{eqnarray}
R(z,\tau)=\frac{1}{\pi\tau}\int d^2\alpha
P(\alpha)\exp\left(-\frac{1}{\tau}|z-\alpha|^2\right).
\label{eqn:Lee}
\end{eqnarray}
The minimum $\tau\in[0,1]$ required for a positive $R$-function is
taken as a nonclassicality measure. Instead of calculating
the integration in Eq.~(\ref{eqn:Lee}), the nonclassical depth $\tau$ for the output state
$|\Psi\rangle\sim(t\hat{a}+r\hat{a}^\dag)|\alpha_0\rangle$ turns out to be unity
regardless of $r$ and $\alpha_0$ by another method: if $\rho$
is orthogonal to a certain coherent state---{\it i.e.},
$\langle\beta|\rho|\beta\rangle=0$ for some $\beta$---the state is
maximally nonclassical, i.e. $\tau=1$ \cite{Barnett}. In our case,
there always exists a certain $\beta^*=-\frac{t}{r}\alpha_0$ such
that
$|\langle\beta|\Psi\rangle|^2\sim|t\alpha_0+r\beta^*|^2|\langle\beta|\alpha_0\rangle|^2=0$.
Therefore, the degree of nonclassicality is maximal if measured by the
nonclassical depth $\tau$ for any
$r\ne0$ of the superposed operation $t\hat{a}+r\hat{a}^\dag$, whereas the nonclassicality increases with $r$ if measured by
the negativity volume $V_N$.

To gain further insight into how the coherent operation produces
nonclassicality, we decompose the output state $\rho_{\rm out}$ into four
parts,
$\rho_{\rm out}\sim|r|^2\rho_{+-}+|t|^2\rho_{-+}+r^*t\rho_{--}+rt^*\rho_{++}$,
where $\rho_{+-}\equiv \hat{a}^\dag\rho_0\hat{a}$ (photon addition), $\rho_{-+}\equiv
\hat{a}\rho_0\hat{a}^\dag$ (photon subtraction), $\rho_{++}\equiv
\hat{a}^\dag\rho_0\hat{a}^\dag$, and $\rho_{--}\equiv
\hat{a}\rho_0\hat{a}$, with $\rho_0$ an initial state. In Fig. 3, we
plot the Wigner function of each part for the input $\alpha_0=0.5$
and $r=\frac{1}{\sqrt {2}}$, where the negative region is colored
in blue. It is well known that the photon subtraction
$\rho_{-+}\equiv \hat{a}\rho_0\hat{a}^\dag$ does not create
nonclassicality at all [Fig. 3 (b)], whereas the photon addition
$\rho_{+-}\equiv \hat{a}^\dag\rho_0\hat{a}$ does [Fig. 3 (a)]. We
see that the ``off-diagonal" components of the operation,
$\rho_{++}+\rho_{--}$ [Fig. 3 (c)], as well as the photon addition [Fig. 3 (a)] affects
the negativity of the whole Wigner distribution [Fig. 3 (d)].

(ii) {\it Thermal states}: A thermal state $\rho_{\rm
th}=(1-e^{-\beta})e^{-\beta a^\dag a}$ with the average photon
number $\bar{n}=\frac{1}{e^\beta-1}$ being as an initial state, we
obtain the Wigner distribution after the coherent operation $t\hat{a}+r\hat{a}^\dag$
as
\begin{eqnarray}
W(\alpha)&=&\frac{1+2\overline{n}}
{(1+\overline{n})(|r|^2+\overline{n})}\left[\frac{\overline{n}}{1+2\overline{n}}|t|^2\right.\nonumber\\
&&-
\frac{(1+2\overline{n})|r|^2+\overline{n}^2}{(1+2\overline{n})^2}
\left(1-\frac{4(1+\overline{n})}{1+2\overline{n}}|\alpha |^2\right)\nonumber\\
&&+4\frac{\overline{n}(1+\overline{n})^2}{(1+2\overline{n})^3}(tr\alpha^2+t^{\ast}
r^{\ast}\alpha^{\ast 2})\left.\right]W_{\rm th}(\alpha),
\end{eqnarray}
where $W_{\rm th}(\alpha)\equiv\frac{2} {\pi
(1+2\overline{n})}e^{-\frac{2|\alpha|^2}{1+2\overline{n}}}$ is the
Wigner distribution of the input thermal state. In Fig. 4, we plot
the Wigner distribution as a function of $r$ for
$\overline{n}=0.1$. On one hand, like the case of coherent-state
input, we see that the negative dip of the Wigner distribution
increases with the ratio $r$ of $\hat{a}^\dag$ in the coherent
operation.
\begin{figure}[tbp]
\centering
\centerline{\scalebox{0.45}{\includegraphics[angle=270]{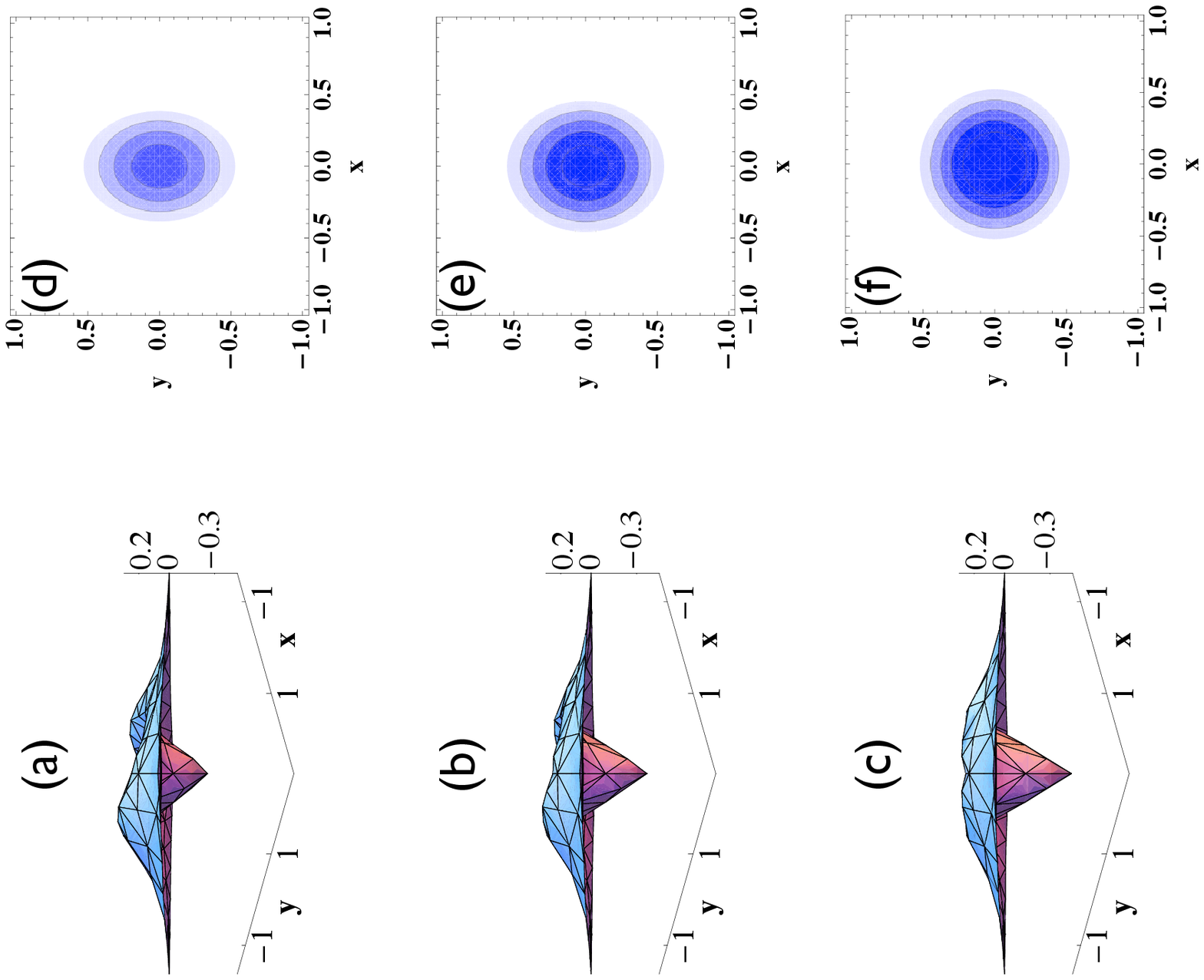}}}
\caption{Wigner distribution $W(\alpha=x+iy)$ after the coherent operation $t\hat{a}+r\hat{a}^\dag$ on a thermal-state input with the average photon number $\overline{n}=0.1$ for (a) $r=1/2$, (b) $r=1/\sqrt{2}$,
 and (c) $r=1$. (d), (e), and (f) are the contour plots corresponding to (a), (b), and (c), respectively, where
 only the negative regions are colored in blue. }
\end{figure}
On the other hand, the negative region appears under the condition
\begin{eqnarray}
C_2x^2+C_3y^2 <C_4, \label{eqn:ellipse}
\end{eqnarray}
where
\begin{eqnarray}
C_2&=&(\overline{n}+r^2+2\overline{n}tr),\nonumber\\
C_3&=&(\overline{n}+r^2-2\overline{n}tr),\nonumber\\
C_4&=&\frac{1+2\overline{n}}{4(1+\overline{n})}
[(1+2\overline{n})r^2-\overline{n}].\nonumber
\end{eqnarray}

Eq.~(\ref{eqn:ellipse}) thus describes an ellipse, as shown in Fig.
4 (d), (e), and (f), with the size of the negative area given by $A_N=\frac{\pi
C_4}{\sqrt{C_2C_3}}$. Note that the negativity emerges only with $r$
above a threshold, $r>\sqrt{\overline{n}/(1+2\overline{n})}$, and
the negative area increases with $r$, unlike the case of
coherent-state input. We also find that the negative volume
increases with $r$, as shown in Fig. 2 (b). In comparison, the degree of
nonclassicality measured by the nonclassical depth turns out to be
$\tau=\frac{(1+\overline{n})r^2}{\overline{n}+r^2}$ using
Eq.~(\ref{eqn:Lee}), which give a
nonzero value for any $r\ne0$ and also increases with $r$.

\section{Observable nonclassical effects}
In this section, we investigate two observable nonclassical effects, quadrature
squeezing and sub-Poissonian statistics, arising from the
coherent operation $t\hat{a}+r\hat{a}^\dag$.

First, the squeezing of a quadrature amplitude
$\hat{X}_{\theta}=\hat{a}e^{-i\theta} +\hat{a}^\dag e^{i\theta}$ is
characterized by $\langle:\Delta^2 \hat{X}_{\theta}:\rangle <0$, where ::
denotes the normal ordering of operators.
On expanding the terms of $\langle:\Delta^2 \hat{X}_{\theta}:\rangle$, one can minimize its value
over the whole angle
$\theta$ \cite{Lee1}, which is then given by
\begin{eqnarray}
S_{\rm opt}&=&\langle:\Delta^2 \hat{X}_{\theta}:\rangle_{\rm min}\nonumber\\
&=&-2|\langle\hat{a}^{\dag
2}\rangle-\langle\hat{a}^\dag\rangle^2|
+2\langle\hat{a}^\dag\hat{a}\rangle-2|\langle\hat{a}^\dag\rangle |^2
,
\label{eqn:S_opt}
\end{eqnarray}
and its negative value in the range of $[-1,0)$ exhibits
nonclassicality. Applying the superposition operation $t\hat{a}+r\hat{a}^\dag$ to a coherent
state $|\alpha_0\rangle$, we obtain
\begin{eqnarray}
S_{\rm opt} = (\frac{2|r|^2}{M}-\frac{1}{2})^2-\frac{1}{4},
\end{eqnarray}
where $M=|t\alpha_0+r\alpha_0^*|^2+|r|^2$, and the maximal degree of squeezing is
thus $-0.25$ under the condition $M=4|r|^2$. In addition, squeezing occurs only under the
threshold condition $M>2|r|^2$.

Let $\alpha_0=|\alpha_0|e^{i\phi}, t=|t|$ and $r=|r|e^{i\phi_r}$. We
obtain the condition $|\alpha_0|=\sqrt{3}|r|/\sqrt{1+2|t||r|\cos(\phi_r-2\phi)}$ for
the optimal squeezing, which in turn gives the optimal $|r|$, given $\alpha_0$, as
\begin{eqnarray}
|r|_{\rm opt}^2=\frac{|\alpha_0|^2(3+2\sqrt{3}|\alpha_0|+2|\alpha_0|^2)}{9+4|\alpha_0|^4}\hspace{0.5cm} (|\alpha_0|\le\sqrt{3}),
\end{eqnarray}
assuming $\cos(\phi_r-2\phi)=1$.
The optimal $|r|_{\rm opt}$ monotonically increases with $|\alpha_0|$,
as shown in the plot of $S_{\rm opt}$ as a function of $|r|$ and $|\alpha_0|$ in Fig. 5.
It implies that the coherent operation $t\hat{a}+r\hat{a}^\dag$ achieves
better squeezing than the mere photon addition
$\hat{a}^\dag$ on a coherent state for $|\alpha_0|\le\sqrt{3}$.
 Moreover, it is known that the photon added state
$\hat{a}^\dag|\alpha_0\rangle$ does not produce squeezing for
$|\alpha_0|<1$ \cite{Agarwal}, whereas the coherent operation always
yields squeezing [Fig.~5]. For $|\alpha_0|\ge\sqrt{3}$, however, the choice of $|r|=1$, i.e. photon addition, becomes optimal for squeezing.
On the other hand, a direct calculation
shows that the superposition operation on a thermal state does not
yield squeezing at all.

\begin{figure}[tbp]
\centering
\includegraphics[width=0.8\columnwidth]{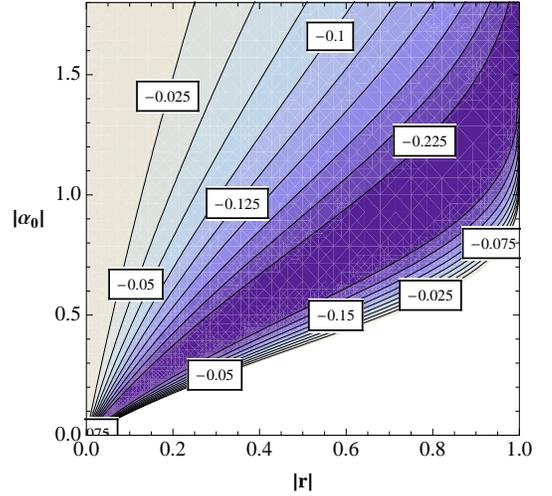}
\caption{Contour plot for $S_{\rm opt}$ in Eq.~(7) of
the state $(t\hat{a}+r\hat{a}^\dag)|\alpha_0\rangle$. Squeezing
occurs in the region above the lowest curve.}
\end{figure}

Second, we consider the sub-Poissonian statistics characterized by
the Mandel $Q$-factor, $Q=\frac{\langle (\Delta
a^{\dag}a)^{2}\rangle} {\langle a^{\dag }a\rangle}-1$.
After the superposition operation on a coherent state, we obtain
\begin{eqnarray}
Q=\frac{|\alpha_0|^2\left[M^2+M-(1-2|r|^2)^2|\alpha_0|^2\right]}{M\left[M+(M-1+2|r|^2)|\alpha_0|^2\right]}-1,
\end{eqnarray}
which always takes a negative value regardless of $\alpha_0,r$, and $t$, with $M=|t\alpha_0+r\alpha_0^*|^2+|r|^2$.
The negativity of Mandel-$Q$ in this case increases with $r$ and decreases with
$\alpha_0$, as shown in Fig. 6 (a). On the other hand, for a
thermal-state input, we obtain
\begin{eqnarray}
Q=\frac{2\overline{n}^2(\overline{n}+r^2)^2-r^4(1+\overline{n})^2}{(\overline{n}+r^2)(2\overline{n}^2+r^2+3\overline{n}r^2)},
\end{eqnarray}
from which we see that the ratio $r$ of $\hat{a}^\dag$ in the
superposition operation must be large enough,
$r^2>\sqrt{2}\overline{n}^2/[1-(\sqrt{2}-1)\overline{n}]$, to
observe sub-Poissonian statistics. The
negativity of Mandel-$Q$ increases with $r$ and decreases with
$\overline{n}$, as shown in Fig. 6 (b).

\begin{figure}[tbp]
\centerline{\scalebox{0.35}{\includegraphics[angle=270]{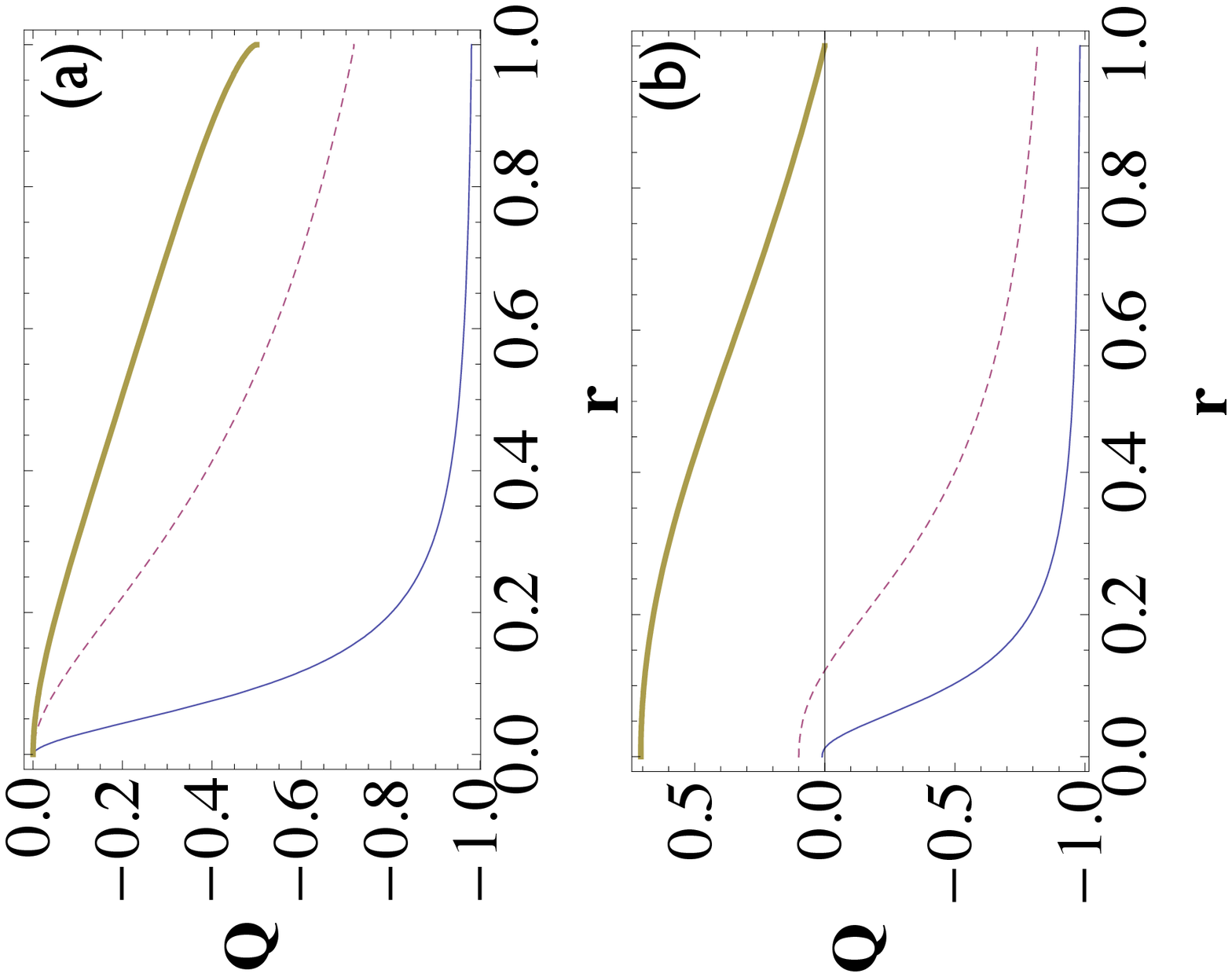}}}
\caption{Mandel Q-factor as a function of $r$ on applying the coherent operation $t\hat{a}+r\hat{a}^\dag$ (a) for a coherent-state input with $\alpha_0=$ 0.1, 0.5, and
1 (from lower to upper curves), and (b) for a thermal-state
input with $\overline{n}=$ 0.01, 0.1, and $1/\sqrt{2}$ (from lower to upper curves). }
\end{figure}

.

\section{Experimental scheme}
In this section, we propose how the superposed operation
$t\hat{a}+r\hat{a}^\dag$ can be implemented in experiment. A key
idea here is to erase which-path information on whether the
implemented operation refers to the photon subtraction $\hat{a}$ or
the photon addition $\hat{a}^\dag$, as described below.

First, note that the success of both the operations, $\hat{a}$ and
$\hat{a}^\dag$, can be heralded by a detection of a single photon in
an optical scheme. When an arbitrary initial state $|\Psi\rangle$ is
injected into a beam-splitter with the other input in a vacuum
state, the detection of a single-photon at one output port heralds
that a photon is subtracted from the initial state, due to the
conservation of photon number. This corresponds to the action
$\hat{a}|\Psi\rangle$, which holds well particularly when the
transmissivity of the beam splitter is large \cite{Wenger}. On the
other hand, if the state $|\Psi\rangle$ is injected to a signal mode
of a nondegenerate parametric amplifier (NDPA) with the idler mode
in a vacuum state, the detection of a single photon at the output
idler mode heralds that one photon is added to the input state, due
to the pairwise photon-creation and -annihilation process in the
NDPA. This corresponds to the action $\hat{a}^\dag|\Psi\rangle$,
which holds well particularly when the interaction strength in the
NDPA is small \cite{Zavatta}. With these two schemes combined, if the
which-path information on the detected single-photon is erased by
using an additional beam splitter ($\rm BS_2$) with transmissivity
$t_2$ [Fig. 7], the coherent superposition $t\hat{a}+r\hat{a}^\dag$
can be conditionally implemented.

In Fig. 7, an arbitrary state $|\Psi\rangle$ is injected into the
parametric down-converter with small coupling strength $s\ll 1$,
which acts as
\begin{eqnarray} \exp(-s\hat{a}^\dag\hat{c}^\dag
+s\hat{a}\hat{c})|\psi\rangle_a |0\rangle_c
\approx (1-s\hat{a}^\dag\hat{c}^\dag)|\psi\rangle_a |0\rangle_c. \nonumber\\
\end{eqnarray}
Next, the BS1 (transmissivity: $t_1\approx1$) acts on the state \cite{Kim1} as
\begin{eqnarray}
&&\hat{B}_{ab}(1-s\hat{a}^\dag\hat{c}^\dag) |\psi\rangle_a |0\rangle_b |0\rangle_c \nonumber\\
&&\approx
(1-\frac{r_1^*}{t_1}\hat{a}\hat{b}^\dag)(1-s\hat{a}^\dag\hat{c}^\dag)
|\psi\rangle_a |0\rangle_b |0\rangle_c,
\end{eqnarray}
Finally, the beam splitter BS2 (transmissivity: $t_2$) with the
transformations $\hat{b}'=t_2\hat{b}+r_2\hat{c}$ and
$\hat{c}'=-r_2^*\hat{b}+t_2^*\hat{c}$ yields
\begin{eqnarray}
&&|\cal {S}_{|\psi\rangle}\rangle\nonumber\\
&&\equiv[1-\frac{r_1^*}{t_1}\hat{a}(t_2\hat{b}^\dag -r_2^{\ast}\hat{c}^\dag)-s\hat{a}^\dag(r_2\hat{b}^\dag +t_2^{\ast}\hat{c}^\dag)\nonumber\\
&&+s\frac{r_1^*}{t_1}\hat{a}\hat{a}^\dag(t_2\hat{b}^\dag
-r_2^{\ast}\hat{c}^\dag)(r_2\hat{b}^\dag +t_2^{\ast}\hat{c}^\dag)]
|\psi\rangle_a |0\rangle_b |0\rangle_c.\nonumber\\
\label{eqn:con_st}
\end{eqnarray}
With the detection of single-photon at PD1 (PD2) and no photon at
PD2 (PD1), we see from Eq.~(\ref{eqn:con_st}) that the state
collapses to $|\psi\rangle_{\rm out}\sim
(t\hat{a}+r\hat{a}^\dag)|\psi\rangle_a$, where
$t\sim\frac{r_1^*}{t_1}t_2$ ($\frac{r_1^*}{t_1}r_2^*$) and $r\sim sr_2$ (
$-st_2^*$). Experimental imperfections, e.g. nonideal
photo-detection, will be further considered in the next section to
investigate the implemented operation particularly on generating an
arbitrary superposition state.

As a remark, we note the identity $\hat{S}^\dag\hat{a}\hat{S}=\hat{a}\cosh s+\hat{a}^\dag e^{i\phi}\sinh s$ and $\hat{S}^\dag\hat{a}^\dag\hat{S}=\hat{a}^\dag\cosh s+\hat{a}e^{-i\phi}\sinh s$, where $\hat{S}\equiv e^{\frac{1}{2}(\xi\hat{a}^{\dag2}-\xi^*\hat{a}^2)}$ is the squeezing operator ($\xi\equiv se^{i\phi}$). 
Thus, one may alternatively implement the coherent operation $t\hat{a}+r\hat{a}^\dag$ by a sequence of squeezing operation, photon subtraction (addition), and the inverse squeezing, $\hat{S}^\dag\hat{a}\hat{S}$ ($\hat{S}^\dag\hat{a}^\dag\hat{S})$, for the case of $|t|>|r|$ ($|t|<|r|$). 
However, this scheme seems to be more demanding than the above proposal based on a single-photon interferometer, as the number of required nonlinear resources is increased. 
Moreover, a very large squeezing $s\rightarrow\infty$ is needed to implement $t\hat{a}+r\hat{a}^\dag$ when the subtraction and the addition parts are comparable to each other, i.e. $|t|\approx|r|$. Even assuming that the squeezing operations can be performed perfectly, we have calculated the output fidelity for superposition states
$C_0|0\rangle+C_1|1\rangle+C_2|2\rangle$ similar to Fig.~8 and found that this alternative scheme does not make improvement. 
Therefore, in the next section, we will focus on the interferometric setting described above to implement the coherent operation.

\begin{figure}[tbp]
\centerline{\scalebox{0.35}{\includegraphics[angle=270]{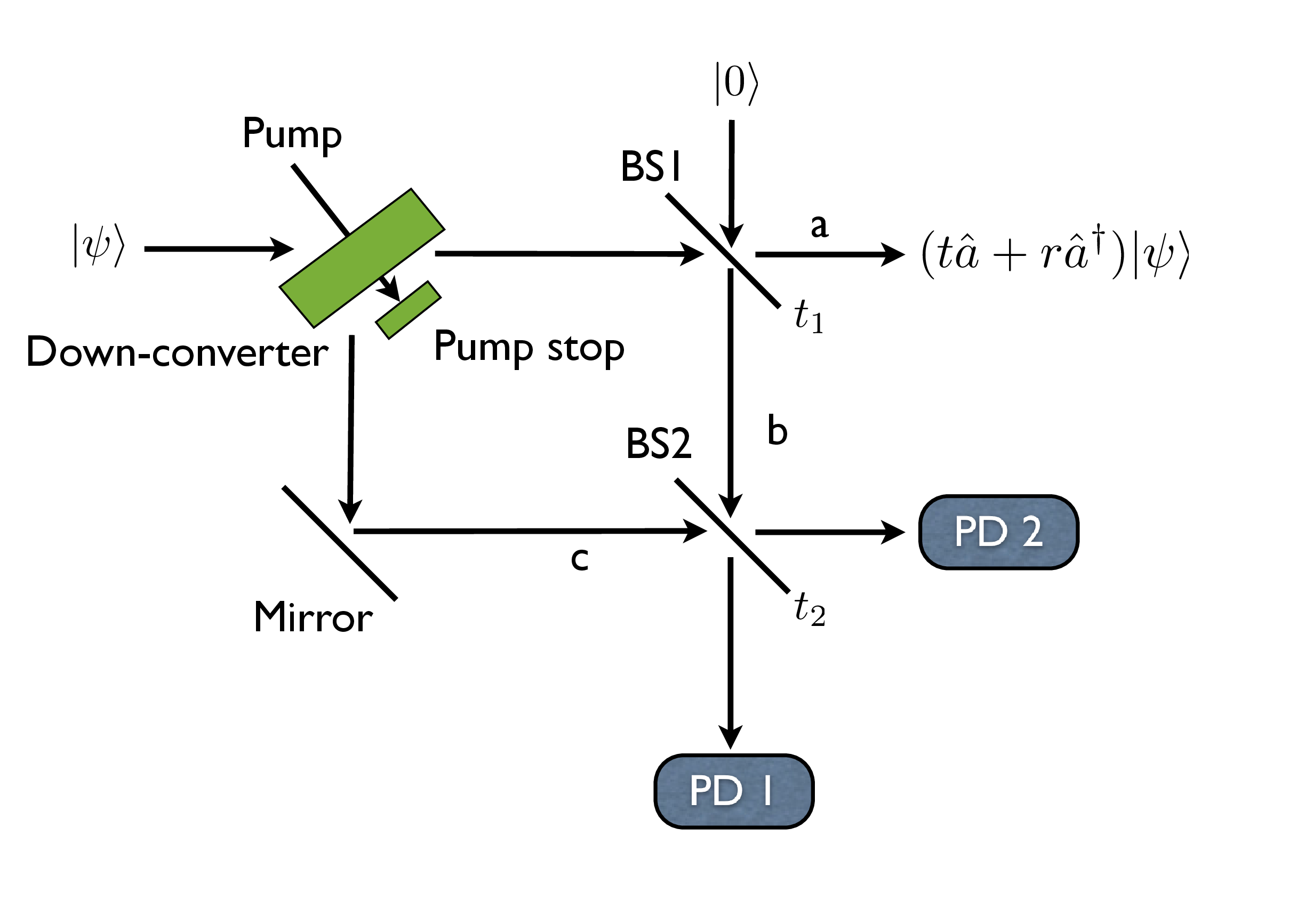}}}
\caption{Experimental scheme to implement the coherent operation
$t\hat{a}+r\hat{a}^\dag$ on an arbitrary state $|\Psi\rangle$. BS1
and BS2 are beam splitters with transmissivitties $t_1$ and $t_2$,
respectively. PD1 and PD2: photo detectors. The coherent operation
is successfully achieved under the detection of a single-photon only
at PD1 or PD2. }
\end{figure}

\section{Generation of a superposition of $|0\rangle$, $|1\rangle$ and $|2\rangle$}
In this section, we show that the coherent operation
$t\hat{a}+r\hat{a}^\dag$ together with a displacement operator $\hat
D(\beta)=e^{\beta\hat{a}^\dag-\beta^*\hat{a}}$ can be employed to
generate an arbitrary field state. In particular, we investigate the
generation of a superposition state involving up to two photons in
detail.

First, note the identity $\hat{O}(\beta,t,r)\equiv \hat
D^\dag(\beta)(t\hat{a}+r\hat{a}^\dag)\hat
D(\beta)=(t\hat{a}+r\hat{a}^\dag+\beta')$, where
$\beta'=t\beta+r\beta^*$, which sequentially represents a displacement, the
coherent operation, and the inverse displacement. If a single photon
state $|1\rangle$ is used as an input, $\hat{O}(\beta,t,r)$ yields a
superposition of number states as
$\hat{O}(\beta,t,r)|1\rangle=t|0\rangle+\beta'|1\rangle+\sqrt{2}r|2\rangle$.
In principle, a succession of $\hat{O}(\beta,t,r)$,
i.e., $\prod_i\hat{O}(\beta_i,t_i,r_i)|1\rangle$ will yield any
desired superposition state by properly choosing the parameters
$\beta_i,t_i$ and $r_i$.

In this paper, we focus on the generation of superposition states
$C_0|0\rangle+C_1|1\rangle+C_2|2\rangle$ with practical
imperfections such as the non-unit photodetection efficiency and the
nonideal single-photon source considered. In particular, we
investigate the effect of on-off detector that does not resolve
photon numbers, which can be represented by a two-component POVM,
$\hat\Pi_0=\sum_n(1-\eta)^n|n\rangle\langle n|$ (no click) and
$\hat\Pi_1=\hat{I}-\hat\Pi_0$ (click), where $\eta$ is the detector
efficiency. On the other hand, a practically generated single-photon
source is identified as a mixture of one-photon and vacuum state,
$\rho_{\rm single}=\eta_s|1\rangle\langle
1|+(1-\eta_s)|0\rangle\langle 0|$, where $\eta_s$ is the source
efficiency \cite{Lvovsky2}. When inserted into the experimental
scheme of Fig.~7 following after the displacement operation $D(\beta)$, the state $\rho_{\rm single}$ is transformed to
$\rho_{\rm coh}=\eta_s|{\cal S}_{|1\rangle_\beta}\rangle\langle {\cal
S}_{| 1\rangle_\beta}|+(1-\eta_s)|{\cal
 S}_{|0\rangle_\beta}\rangle\langle {\cal S}_{|0\rangle_\beta}|$, where $|\cal
{S}_{|\psi\rangle}\rangle$ is given in Eq.~(\ref{eqn:con_st}) and $|n\rangle_\beta\equiv {\hat D}(\beta)|n\rangle$.
Thus, the output state emerges after the inverse displacement ${\hat D}^\dag(\beta)$, under the condition of click at PD1 and no-click at PD2, as $\rho_{\rm out}=\frac{{\rm
Tr}_{b,c}\left[\rho_{\rm con}\right]}{{\rm
Tr}_{a,b,c}\left[\rho_{\rm con}\right]}$, where
$\rho_{\rm con}\equiv \hat{D}_a^\dag(\beta)\rho_{\rm
coh}\hat{D}_a(\beta)\cdot\hat\Pi_1^b\otimes\hat\Pi_0^c$.
Calculation shows
\begin{eqnarray}
\rho_{\rm con}=&&\eta_s\eta\left(|\Phi\rangle\langle\Phi|+{\cal A}|\phi\rangle\langle\phi|\right)\nonumber\\
&&+(1-\eta_s)\eta\left(|\Psi_1\rangle\langle\Psi_1|+{\cal A}|\Psi_2\rangle\langle\Psi_2|\right),
\end{eqnarray}
where
\begin{eqnarray}
|\Phi\rangle&=&R_1t_2|0\rangle+(R_1t_2\beta+sr_2\beta^*)|1\rangle+\sqrt{2}sr_2|2\rangle,\nonumber\\
|\phi\rangle&=&\beta^*|0\rangle+(2+|\beta|^2)|1\rangle+\sqrt{2}\beta|2\rangle,\nonumber\\
|\Psi_1\rangle&=&(R_1t_2\beta+sr_2\beta^*)|0\rangle+sr_2|1\rangle,\nonumber\\
|\Psi_2\rangle&=&(1+|\beta|^2)|0\rangle+\beta|1\rangle,\nonumber\\
{\cal A}&=&s^2|R_1|^2(1-\eta+2\eta|t_2r_2|^2)
\label{eqn:prac_st}
\end{eqnarray}
with $R_1=\frac{r_1^*}{t_1}$.

\begin{figure}[tbp]
\centering
\includegraphics[width=0.8\columnwidth]{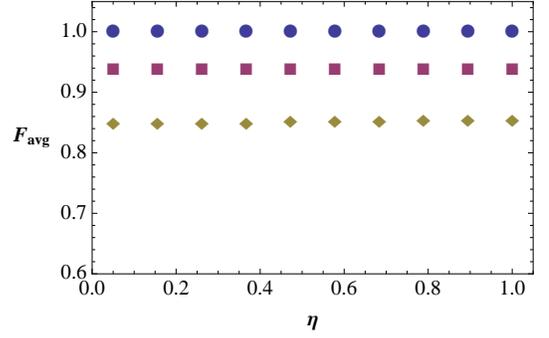}
\caption{Average fidelity $F_{\rm avg}$ between an ideal superposition state
and an experimentally realizable state as a function of detector efficiency $\eta$ (on-off detector) for the single-photon source efficiency
$\eta_s=0.69$(diamond), $0.85$(square), and $1$(circle), with $s=R_1=0.01$. }
\end{figure}

To evaluate the overall performance of the proposed scheme, we calculate the average fidelity of the produced state $\rho_{\rm out}$ with a target state
$|\Phi\rangle=\sin{\theta}\cos{\phi}|0\rangle+\sin{\theta}\sin{\phi}|1\rangle
+\cos{\theta}|2\rangle$, that is, $F_{\rm avg}=\frac{1}{4\pi}\int^{\pi}_{0}d\theta \sin{\theta}\int^{2\pi}_{0}d\phi
\langle \Phi |\rho_{\rm out}^{\{\theta, \phi\}}|\Phi\rangle$ over the entire range of angles $\theta$ and $\phi$.
Note that $|\Phi\rangle$ in Eq.~(\ref{eqn:prac_st}) is the ideal state that would be obtained using a perfect single-photon source ($\eta_s=1$) and a perfect photon-number-resolving detector. Thus, by identifying it with the state $|\Phi\rangle=\sin{\theta}\cos{\phi}|0\rangle+\sin{\theta}\sin{\phi}|1\rangle
+\cos{\theta}|2\rangle$, we obtain the state parametrization as
 \begin{eqnarray}
&&t_2=\frac{\sqrt{2}sB_1}{\sqrt{R_1^2+2s^2B_1^2}},~
r_2=\frac{|R_1|}{\sqrt{R_1^2+2s^2B_1^2}},\nonumber\\
&&\beta=\frac{\sqrt{2}B_2}{1+\sqrt{2}B_1},
\end{eqnarray}
where $B_1\equiv \tan{\theta}\cos{\phi}$ and $B_2\equiv \tan{\theta}\sin{\phi}$.

In Fig. 8, the average fidelity $F_{\rm avg}$ is plotted as a function of the detector efficiency $\eta$ for various source efficiencies $\eta_s=0.69, 0.85$, and 1, with $s=R_1=0.01$.
From Eq.~(14), we see that even with a perfect single-photon source ($\eta_s=1$) and a unit detector efficiency ($\eta=1$), the contribution of the unwanted state $|\phi\rangle$ to the fidelity $F_{\rm avg}$ cannot be completely eliminated [Cf. the coefficient $\cal A$ in Eq.~(\ref{eqn:prac_st})]. 
This is due to the use of on-off detector that does not distinguish photon-numbers and $|\phi\rangle$ arises in the output state $\rho_{\rm con}$ due to the detection of higher-number of photons than a single-photon.
In general, nevertheless, the fidelity $F_{\rm avg}$ is largely insensitive to the detector efficiency $\eta$, and depends substantially on the single-photon source efficiency $\eta_s$ only.
With $\eta_s=0.69$ previously reported in \cite{Lvovsky2}, a high fidelity $F_{\rm avg}\sim0.85$ seems to be readily achievable within the current technology.

\section{Conclusion}
In this paper, we have investigated a coherent superposition $t\hat{a}+r\hat{a}^\dag$ acting on continuous variable systems
that can be used for quantum state engineering. It has been shown that the operation creates nonclassicality out of a classical state with observable effects such as squeezing and sub-Poissonian statistics. For the coherent-state inputs, the degree of nonclassicality measured by negative volume in phase space increases with the ratio $r$, while that measured by nonclassical depth due to C. T. Lee is maximal regardless of $r$. In particular, given an input coherent amplitude $\alpha_0$, the squeezing effect occurs optimally for $r<1$, which implies that the coherent operation generally achieves better squeezing than the bare photon addition $\hat{a}^\dag$.
For the thermal-state inputs, the degree of nonclassicality increases with $r$ measured by both the negative volume and the nonclassical depth.

We have also proposed an optical experimental scheme to implement the coherent operation $t\hat{a}+r\hat{a}^\dag$ in a single-photon intereference setting.
Furthermore, it has been shown that the coherent operation combined with the displacement operations can be employed to generate an arbitrary superposition state and that a high fidelity particularly in producing a superposition state involving up to two photons is achievable against experimental imperfections using the currently available techniques.

\begin{acknowledgments}
This work is supported by the NPRP grant 1-7-7-6 from Qatar National Research Fund.

\end{acknowledgments}

\end{document}